\documentclass[a4paper,fleqn,usenatbib]{mn2e}

\usepackage[T1]{fontenc}
\usepackage{ae,aecompl}
\usepackage{xspace}
\usepackage{graphicx}	% Including figure files
\usepackage{amsmath}	% Advanced maths commands
\usepackage{amssymb}	% Extra maths symbols
\def\msun{M_\odot}

\newcommand{\eg}{{e.g.},\,}
\newcommand{\ie}{{i.e.},\,}
\newcommand{\Heii}{\ion{He}{II}\xspace}
\newcommand\ion[2]{\text{#1\,\textsc{\lowercase{#2}}}}	% ionization states 
\newcommand\apss{Ap\&SS}             % Astrophysics and Space Science

\title[OGLE16aaa - Hungry Super Massive Black Hole]{OGLE16aaa - a Signature of a Hungry Super Massive Black Hole}

\author[{\L}. Wyrzykowski et al.]{
{\L}ukasz Wyrzykowski,$^{1}$\thanks{E-mail: lw@astrouw.edu.pl, name pronunciation: {\it Woocash Vizhikovski} }
M.Zieli{\'n}ski,$^{1}$
Z.Kostrzewa-Rutkowska,$^{1,2,3}$
A.Hamanowicz,$^{1}$
\newauthor
P.G.~Jonker,$^{2,3}$
I.Arcavi,$^{4,5}$
J.Guillochon,$^{6}$
P.J.~Brown,$^{7}$
S.Koz{\l}owski,$^{1}$
A.Udalski,$^{1}$
\newauthor
M.K.Szyma{\'n}ski,$^{1}$
I.Soszy{\'n}ski,$^{1}$
R.Poleski,$^{1,8}$
P.Pietrukowicz,$^{1}$
J.Skowron,$^{1}$
P.Mr{\'o}z,$^{1}$\newauthor
K.Ulaczyk,$^{1,9}$
M.Pawlak,$^{1}$
K.A.Rybicki,$^{1}$
J.Greiner,$^{10}$
T.Kr{\"u}hler,$^{10}$
J.Bolmer,$^{10,11}$\newauthor
S.J.Smartt,$^{12}$
K.Maguire,$^{12}$ 
K.Smith$^{12}$
\\
$^{1}$Warsaw University Astronomical Observatory, Al. Ujazdowskie 4, 00-478 Warszawa, Poland\\
$^{2}$SRON, Netherlands Institute for Space Research, Sorbonnelaan 2, 3584 CA Utrecht, the Netherlands\\
$^{3}$Department of Astrophysics/IMAPP, Radboud University Nijmegen, P.O. Box 9010, 6500 GL Nijmegen, the Netherlands\\
$^{4}$Las Cumbres Observatory Global Telescope Network, 6740 Cortona Dr Ste 102, Goleta, CA 93117-5575, USA\\
$^{5}$Kavli Institute for Theoretical Physics, University of California, Santa Barbara, CA 93106-4030, USA\\
$^{6}$Harvard-Smithsonian Center for Astrophysics, 60 Garden St., Cambridge, MA 02138, USA\\
$^{7}$George P. and Cynthia Woods Mitchell Institute for Fundamental Physics \& Astronomy, Texas A. \& M. University,\\~~~Department of Physics and Astronomy, 4242 TAMU, College Station, TX 77843, USA\\
$^{8}$Ohio State University, Department of Astronomy, 140 West 18th Avenue, Columbus, OH, USA 43210\\
$^{9}$Department of Physics, University of Warwick, Gibbet Hill Road, Coventry, CV4 7AL, UK\\
$^{10}$Max-Planck Institute for Extraterrestrial Physics, 85748 Garching, Giessenbachstr.1, Germany\\
$^{11}$European Southern Observatory, 85748 Garching, Germany\\
$^{12}$Astrophysics Research Centre, School of Mathematics and Physics, Queens University Belfast, Belfast BT7 1NN, UK
}

\date{Accepted after long and painful review process. Received long time ago; in original form ZZZ}

\pubyear{2016}

\begin{document}
\label{firstpage}
\pagerange{\pageref{firstpage}--\pageref{lastpage}}
\maketitle

\begin{abstract}
We present the discovery and first three months of follow-up observations of a currently on-going unusual transient detected by the OGLE-IV survey, located in the centre of a galaxy at redshift z=0.1655. 
The long rise to absolute magnitude of -20.5 mag, slow decline, very broad He and H spectral features make OGLE16aaa similar to other optical/UV Tidal Disruption Events (TDEs). 
Weak narrow emission lines in the spectrum and archival photometric observations suggest the host galaxy is a weak-line Active Galactic Nucleus (AGN), which has been accreting at higher rate in the past. 
OGLE16aaa, along with SDSS J0748, seems to form a sub-class of TDEs by weakly or recently active super-massive black holes (SMBHs). 
This class might bridge the TDEs by quiescent SMBHs and flares observed as ``changing-look QSOs'', if we interpret the latter as TDEs. 
If this picture is true, the previously applied requirement for identifying a flare as a TDE that it had to come from an inactive nucleus, could be leading to observational bias in TDE selection, thus affecting TDE-rate estimations. 
\end{abstract}


%%%%%%%%%%%%%%%%%%%%%%%%%%%%%%%%%%%%%%%%%%%%%%%%%%

%%%%%%%%%%%%%%%%% BODY OF PAPER %%%%%%%%%%%%%%%%%%
\section{Introduction}
It has become a paradigm that nearly all galaxies at current times harbour a super-massive black hole (SMBH) in their centre (\eg \citealt{1998AJ....115.2285M}). In the cold dark matter ($\lambda$CDM) theory of cosmology, current (redshift zero) galaxies are the product of hierarchical mergers of smaller galaxies. These smaller building blocks also host black holes in their centres (\citealt{1995ARA&A..33..581K}, \citealt{2012NatCo...3E1304G}), potentially intermediate-mass black holes (IMBHs), with masses from 100 to 10,000$\msun$. After two galaxies merge, the two black holes will merge as well (see \citealt{1980Natur.287..307B}). Therefore, mergers of black holes may play an important role in building SMBHs (cf. \citealt{2007ApJ...665..107P}). Interestingly, SMBHs with masses of more than a billion $\msun$ have been found already at redshifts of more than 6 when the Universe was less than 1 Gyr old (see \citealt{2006NewAR..50..665F}). Such SMBHs may be seeded by 100 
$\msun$ black holes at redshifts z$>$15 which then grow by uninterrupted accretion of gas at the Eddington rate with a standard radiative efficiency of 10  per cent (\eg \citealt{2013ASSL..396..293H}). However, quasars grow only for about $\approx$4.5$\times$10$^7$ years before feedback stops the gas supply \citep{1998A&A...331L...1S}. In order to solve this problem one can start with more massive black holes such as IMBHs and/or allow mass to be accreted at a rate higher than the Eddington limit and/or have part of the black hole growth be due to mergers of black holes.

Tidal Disruption Events (TDEs, \eg \citealt{Hills1975}, \citealt{Rees1988}), in which a star is torn apart by the tidal forces of the SMBH, offer a unique opportunity to study the mass distribution of SMBHs.  The intrinsic TDE rate should be dominated by the SMBHs with the lowest mass (\citealt{2004ApJ...600..149W}, \citealt{StoneMetzger2016}), so volume--complete TDE samples can
measure the occupation fraction of IMBHs in small galactic bulges informing SMBH formation theories. However, the inhomogeneous and small sample of TDEs 
currently available, found either in X--rays (\eg \citealt{Bade1996}, \citealt{2013A&A...552A..75N}) or in the UV/optical (\eg \citealt{vanVelzen2011}, \citealt{Gezari2012}, \citealt{Arcavi2014}, \citealt{Holoien2014}) 
prevents us from discriminating between various emission mechanisms of TDEs and hence from conclusions on the SMBH mass function (\eg \citealt{StoneMetzger2016}). % IA: Suggest adding here also references to Piran et al. 2015 and Krolik et al. 2016 as they exemplefy a lot of the current debate. -> LW: would like to,but the text is already too long!

Optical/UV TDEs are relatively luminous ($M_{peak}\sim-20$) and blue ($T\sim$few $10^4$K) few-months- to years-long transients with broad H and/or \Heii emission lines\citep{Arcavi2014}. 
%It is not yet clear if and how the optical/UV events relate to the variety of X-ray and $\gamma$-ray flares also proposed as TDEs. %For a recent review on TDEs we refer the reader to \citet{Komossa2015}. 
% IA: There's also a recent Gezari review, should cite that as well. -> LW: couldn't find it. Just killing the Komossa review reference.
\cite{Arcavi2014} and \cite{FrenchArcavi2016} noted that 75 per cent of optically found TDEs occurred in quiescent Balmer-strong (called E+A by some) % IA: Replaced E+A with "quiescent Balmer-strong" (more accurate / less controversial definition).-> sure, but you used E+A in your papers, so for completness of nomenclature I'd like to use it, at least here.
galaxies, which account for $2.3$ per cent of SDSS galaxies. Such galaxies are 
thought to be products of a recent merger (within 1~Gyr), which triggered an observed increase in star formation (Balmer absorption series in their spectra are caused by a significant amount of A--type stars). That preference might be due to disturbed dynamics of the nuclear star cluster, or the presence of a coalescing binary black hole, causing nearby stars to go on a collision course with their central black holes (e.g.~\citealt{2011ApJ...738L...8W}).

However, a fraction of the optical/UV TDEs seems to not match such scenario. 
Host galaxy spectra of Extreme Coronal Line Emitters (ECLE), SDSS J095209.56+214313.3 \citep{Komossa2008, Palaversa16} and SDSS J0748 \citep{Wang2011}, as well as  ASASSN-14li \citep{2016MNRAS.455.2918H} and PTF09axc \citep{Arcavi2014} 
present weak, narrow emission lines, which could be indicative of a weak Active Galactic Nucleus (AGN) present in the core.

AGNs are known to exhibit photometric variability at a level of few tens of magnitude (\eg~\citealt{Kozlowski2016}). 
However, occasionally, flares are observed well above the level of their typical variability, both in X-rays (\eg \citealt{Strotjohann2016}) and optical (\eg \citealt{Tanaka2013}). 
The reasons for such significant changes in mass accretion rate are still under debate and include binary black hole interactions with the disk as well as stellar disruptions. 
%Variability and transient activity was also detected in case of narrow-line emitters (LINERs), \eg \cite{Maoz2005}.

%%%% CHANGING LOOK QSOs and Bin BHS
Long-term and wide-field spectroscopic and photometric data obtained by the Sloan Digital Sky Survey (SDSS), as well as on-going transient searches (ASAS-SN, Pan-STARRS, MASTER surveys), have revealed nearby AGNs (Seyfert galaxies), that changed their spectral characteristics, often accompanied with a temporal increase in observed flux, e.g., \cite{Lawrence2016}. 
In particular, an event observed in the quasar SDSS~J0159$+$0033 \citep{2015ApJ...800..144L}, has been recently interpreted as a TDE by a massive SMBH \citep{Merloni2015}. 
In large fraction of other similar object, so called, ``changing-look quasars'' (hereafter, CH-L-QSOs) \citep{Macleod15}, 
%
%have shown that they generally follow the following pattern: a broad H$\alpha$ emission line appears on top of a narrow one and this is accompanied with a significant rise in the blue continuum.
%In a large fraction of these objects, 
the appearance of a broad component to the H$\alpha$ emission line was found to be transient \citep{Baldassare16}, \ie the broad line disappeared on the time-scale of years. 
Moreover, one of such CH-L-QSOs, found in NGC 2617 by ASAS-SN in 2013 \citep{Shappee14} 
has recently been reported to show a re-brightening (\citealt{Oknyansky16})
.
If those flares are indeed due to TDEs, the stellar disruption must be induced fairly frequently, namely, on a time-scale of years.
An X-ray TDE candidate in an AGN IC~3599 was found to show a re-occurrence period of about 9 years\citep{Campana15}, which could be explained with a binary central black hole.
% (\eg \citealt{Ricarte2016}). 
%Another candidate for repeating TDEs has been published by \cite{Campana15}, who found repeated TDE-like X-ray flares related to an active nucleus in the IC~3599 galaxy, with re-occurrence period of about 9 years.
%Explanations of the high TDE rate necessary to explain multiple TDEs within one decade, include mergers of galaxies (\eg \citealt{FrenchArcavi2016}), resulting in, either dynamical instabilities in the central stellar cloud of the main galaxy, or a binary central black hole, which disturbs stellar orbits (\eg \citealt{2011ApJ...738L...8W}; \citealt{Liu13}; \citealt{Ricarte2016}).

In this Letter we describe the discovery and first months of follow-up of an unusual transient, OGLE16aaa. We propose that OGLE16aaa is a TDE candidate in a low--luminosity AGN host galaxy and it may be another link in chain of disruption events that occur from quiescent to active galaxy cores.

% FIG: CENTROID %%%
\begin{figure}
	\centering
	\includegraphics[width=1\columnwidth]{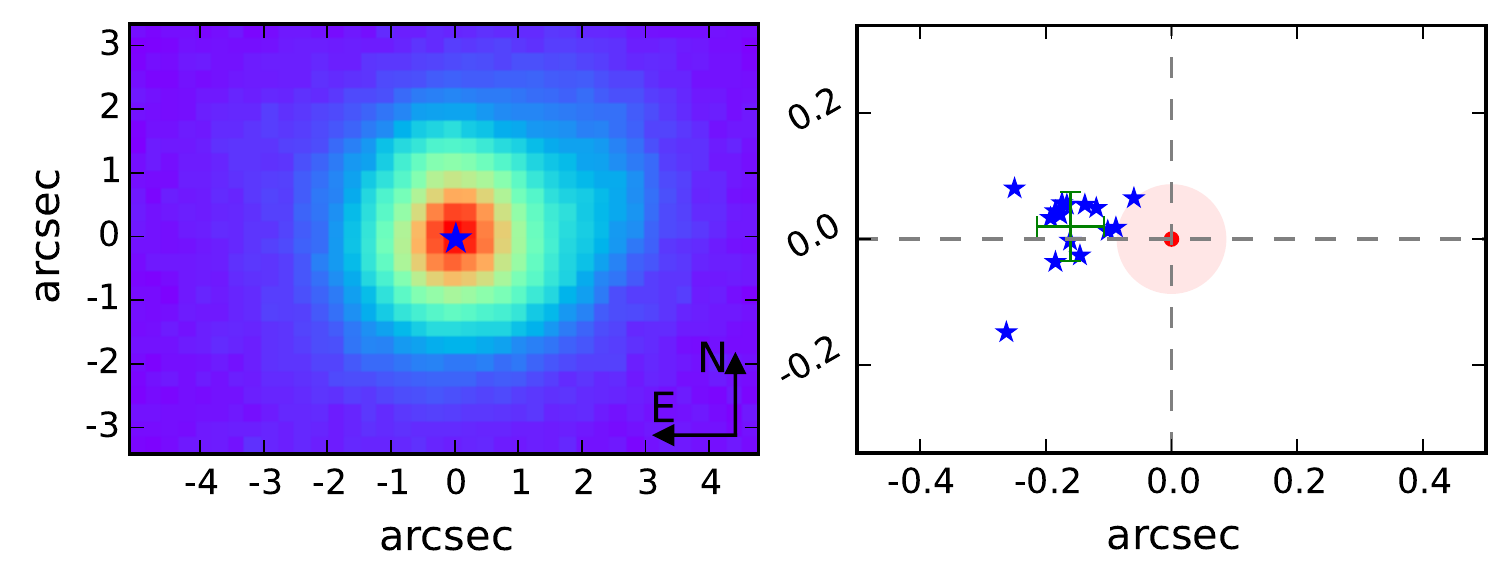}
    \caption{Left: mean position of the OGLE16aaa transient (asterisk) on false colour deep stack of $\sim$100 pre-flare OGLE $I$-band images. Right: positions of the detections when the source was brightest (blue) with respect to the centre of the galaxy (red circle). Green cross marks the mean position of the transient offset from the centre by 160$\pm$140 mas, when including the error-bar on the best-fit galaxy position.}
\label{fig:chart}
\end{figure}

%%%%%%%%%%%%%%%%%%%%%%%%%%%%%%%%%%%%%%%%%%%%%%%%%%
\section{Discovery and early follow-up}
\begin{figure}
	\includegraphics[width=1\columnwidth]{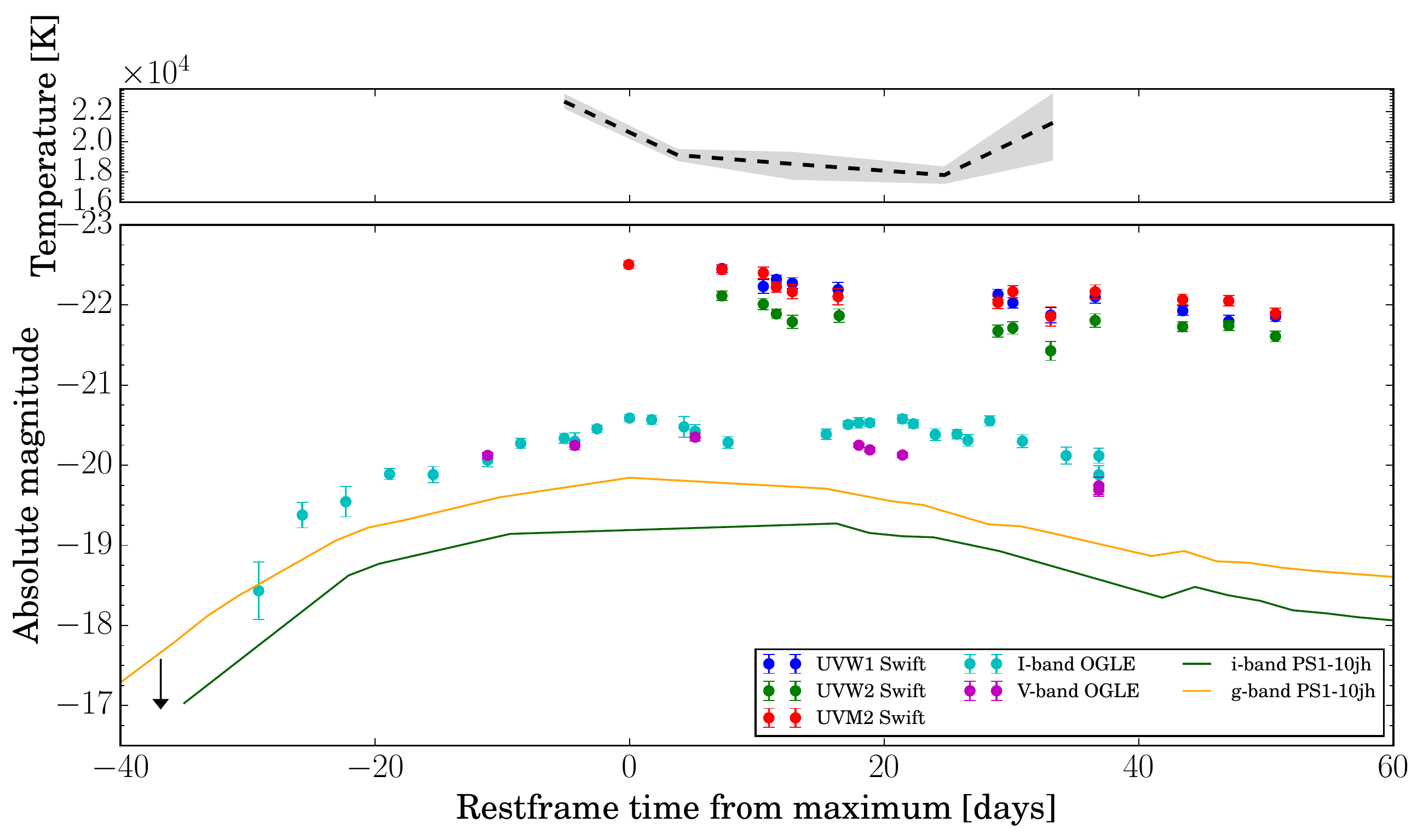}
    \caption{Photometric observations (bottom) and temperature (top) evolution with 1-$\sigma$ uncertainty (grey) of the OGLE16aaa transient. Also shown is the light curve of the TDE PS1-10jh\citep{Gezari2012}. An arrow shows the last non-detection in OGLE.}
\label{fig:lc}
\end{figure}

OGLE16aaa (Fig. \ref{fig:lc}) was discovered by the OGLE Transient Detection System \citep{2014AcA....64..197W}
\footnote{http://ogle.astrouw.edu.pl/ogle4/transients}
, a programme within the Optical Gravitational Lensing Experiment (OGLE-IV, \citealt{2015AcA....65....1U}).
%, which finds on-going transients in nearly 700 sq.deg. of sky around the Magellanic Clouds since 2012. 
This transient, located at RA, Dec(J2000.0) = 1:07:20.88, -64:16:20.7, was found during visual inspection of candidates detected by the automated pipeline on January 2nd, 2016 %(Heliocentric Julian Date; HJD-2450000=7389.55822) 
at I-band magnitude $\sim$20 mag \citep{2016ATel.8577....1W}. 
It was found at the centre of GALEXASC J010720.81-641621.4 galaxy of 17.00$\pm$0.01 mag as measured by OGLE in the I-band and 
20.83$\pm$0.12 mag and 21.82$\pm$0.32 mag in GALEX NUV and FUV, respectively. 
Fig.~\ref{fig:chart} shows the host galaxy image from a deep stack of $\sim$100 OGLE I-band images (all taken before the event) with the position of the transient.
% measured using Difference Image Analysis (DIA,\citealt{2015AcA....65....1U}).  

% FIG: LIGHT CURVE %%%
%%%%

\begin{figure}
	\includegraphics[width=1\columnwidth]{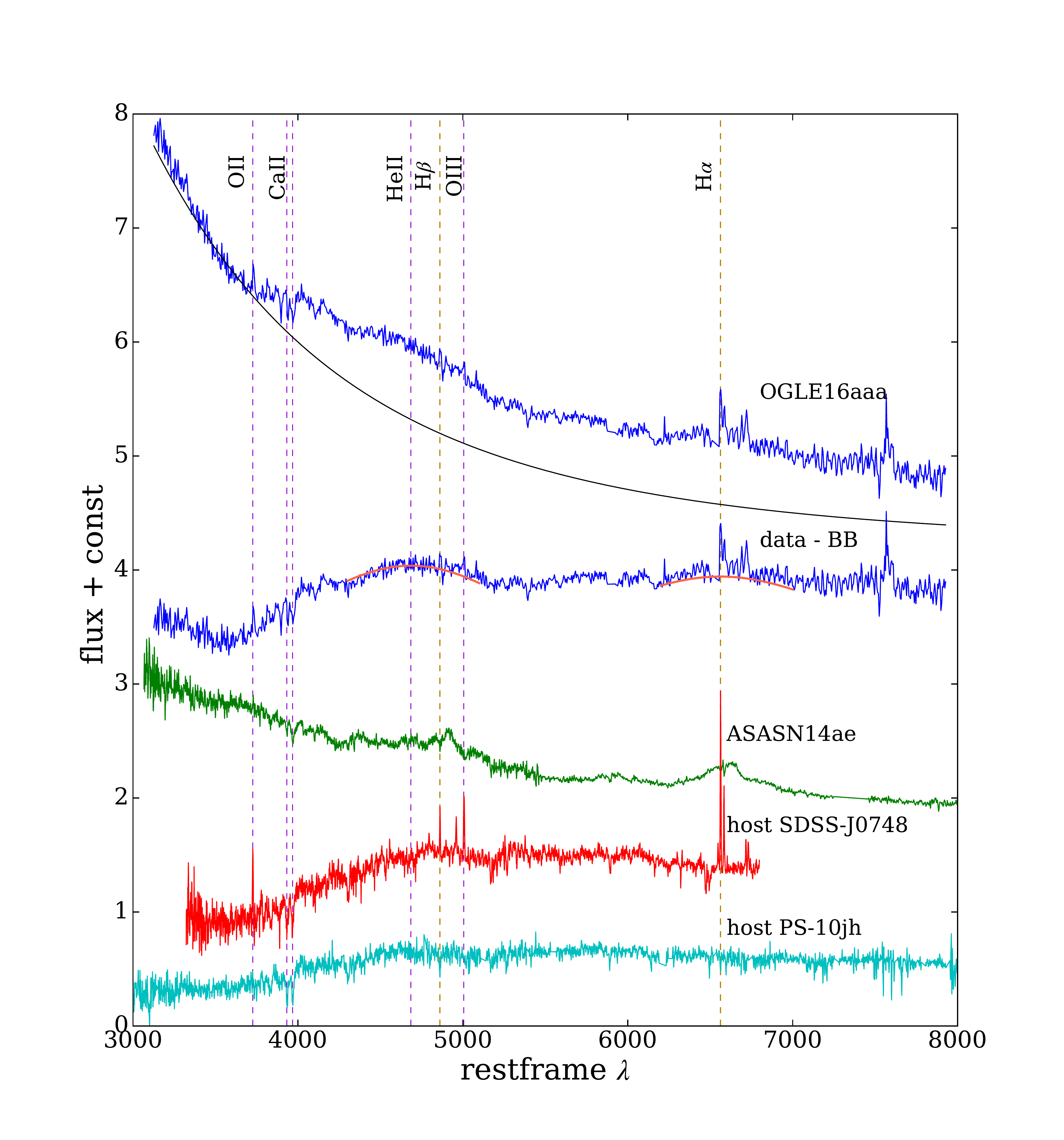}
    \caption{Top: PESSTO spectrum of OGLE16aaa flare (with the Gr\#13 and 1\farcs0 slit setup) at -3 days of the $I$-band maximum. The solid line shows the black body model with T=22,000~K best matching to the blue part of the spectrum. Below is the same spectrum, but with BB model subtracted. Orange Gaussians indicate the location of broad spectra components at \Heii and H$\alpha$. 
  We also show spectra of TDE-candidate hosts with narrow emission lines. Also shown are the spectra of hosts of other TDE candidates: SDSS J0748 and PS-10jh.}
\label{fig:spec}
\end{figure}

On 17 Jan 2016 the Public ESO Spectroscopic Survey for Transients\footnote{www.pessto.org} (PESSTO, \citealt{2015A&A...579A..40S}) took a spectrum of OGLE16aaa with the ESO/NTT EFOSC2 instrument \citep{2016ATel.8559....1F} (top of Fig.~\ref{fig:spec}). The spectrum (which is available on $WISEREP$ \citealt{WISEREP}) shows 
a blue continuum with broad \Heii and H$\alpha$ and weak emission and absorption lines consistent with a redshift of z=0.1655 ($\sim$800Mpc). The light curve reached an $I$--band maximum on 2016-01-20.1 at 18.98$\pm$0.05 mag. 
Assuming standard cosmology ($\Omega_{\rm M}$=0.28, $H_0$=70), and taking into account only the Galactic extinction towards this source of $A_I=0.028$ mag \citep{2011ApJ...737..103S}, the maximum brightness corresponds to an absolute magnitude of $-$20.5 mag, in range consistent with previous TDEs \citep{Arcavi2014}.

The transient was then followed-up from the ground with the GROND instrument on the MPE 2.2m telescope in La Silla \citep{2008PASP..120..405G} and with {\it Swift} satellite's Ultra-Violet Optical Telescope (UVOT) from 2016-01-19 (\citealt{2016ATel.8579....1G}, \citealt{2016ATel.8644....1Z}). For details on {\it Swift} data reductions see \citet{Brown2014}.

Difference imaging photometry conducted by OGLE on the densely sampled pre-discovery images spanning about 3.5 years showed no prior flaring nor variable activities in the core of this galaxy at a level below 1 per cent. 
%(see Fig. \ref{fig:history})
The last non-detection on HJD 2457364.61144 and first detection on 2457373.5795 at 21.1$\pm$0.4 mag, allow us to constrain the beginning of the event to about HJD=2457369$\pm$4 days (see Fig. \ref{fig:lc}).

% % FIG: HISTORY %%%
% \begin{figure}
% 	\includegraphics[width=\columnwidth]{figure-history.pdf}
%     \caption{Historical light curve prior to the flare covering 3.5 years obtained with difference imaging at the centre of the galaxy. Red arrows indicate non-detections or data fainter than 22 mag. Individual detections prior to the flare are all below 21 mag and are caused by fluctuations in the image subtractions and seeing variations.
%     }
% \label{fig:history}
% \end{figure}
% %%%%

% FIG: SPEC %%%
%%%%

%FIG: TDEFit model from James
\begin{figure}
	\centering\includegraphics[width=1\columnwidth]{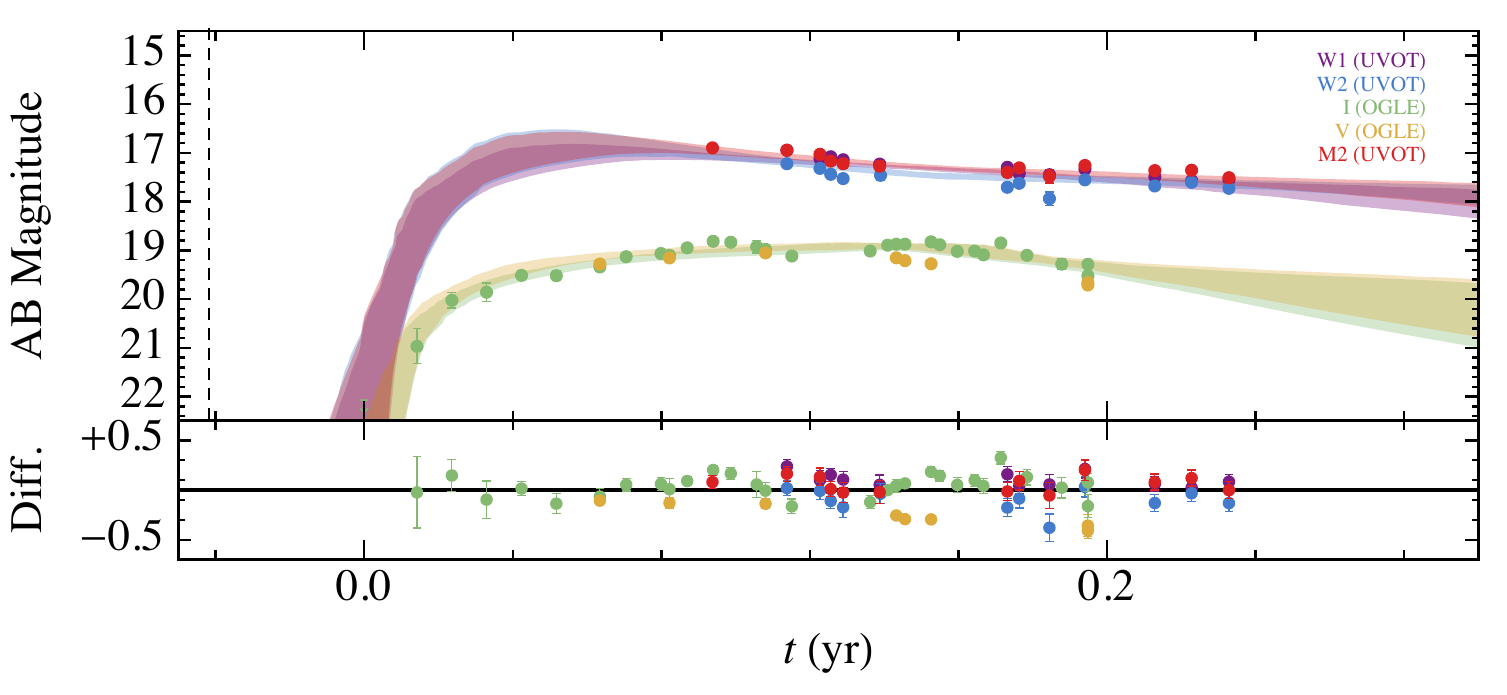}
    \caption{{\tt TDEFit} model of the OGLE16aaa transient fit to OGLE and {\it Swift} data. The shaded regions in the upper panel show the 1-$\sigma$ spread in the model fits, whereas the lower panel shows residuals of the highest-scoring model.}
\label{fig:tdemodel}
\end{figure}
%%%%

%%%%%%%%%%%%%%%%%%%%%%%%%%%%%%%%%%%%%%%%%%%%%%%%%%
\section{Host and transient characteristics}
%
%\subsection*{Location}
\textbf{Location}. 
%The position of the centre of the host was derived from OGLE $I$-band images from around maximum using DIA with a typical accuracy better than half an OGLE-IV pixel (0.26'' per pixel, see \citealt{2014AcA....64..197W}). 
The offset of the transient from the galaxy's photo-centre, obtained from the OGLE $I$-band images using DIA, is less than 160$\pm$140 mas (460$\pm$400 pc for z=0.1655), as shown in Fig. \ref{fig:chart}, and is consistent with the position of the nucleus.

%%%% HOST GALAXY
\noindent
\textbf{Host galaxy}. 
From the light distribution of the host on the deep OGLE $I$-band image we obtained a Sersi{\'c} index of n=1.08, which corresponds to a black hole mass of $\log M_h =6.58$ ($\approx4\times10^6~\msun$) using the relation in \citep{2016ApJ...821...88S}.

Since there is, to our knowledge, no pre-flare host spectrum available, we have to wait until the event fades out completely to obtain the spectrum of the nucleus of the galaxy alone. 
However, the La PHARE\citep{Ilbert2006} best matching template spectrum to the spectral energy distribution (SED) from archival UV, optical, near- and far- infrared 
 %(GALEX: NUV, FUV), optical (OGLE: I), NIR (2MASS: J, H, K) and IR (WISE: W1,W2,W3,W4) 
 observations is found for a galaxy with a stellar mass of $\log M = 10.3\pm0.2 \msun$, with no strong evidence for star formation.
 % and resembling a weak AGN\citep{Assef2010}.

Signatures of the host are already present in the PESSTO spectrum of the flare.
The spectrum shows weak [OIII], [OII] and [NII] narrow forbidden lines in emission as well as narrow emission of $H\alpha$ and $H\beta$. 
Assuming the broad features are from the transient and the narrow ones are from the host, the host spectrum 
resembles that of SDSS J0748 \citep{Wang2011}
which hosted a TDE candidate. 
The galaxy does not seem to belong to the quiescent Balmer-strong galaxies,
which were noted to host most of the optically discovered TDEs so far \citep{FrenchArcavi2016}.
The line ratios (log($[$NII]/H$\alpha)=-0.43$ $vs$ log($[$OIII]/H$\beta)=-0.25$)
indicate there is a mixture of star formation in the galaxy of
OGLE16aaa as well as a weak AGN in its core. 
The line ratios are also similar to those in the host of SDSS J0748. 
Moreover, the WISE\citep{Wright2010} colours for OGLE16aaa and SDSS J0748 place those hosts near the region occupied by AGNs, following the classification method of \cite{Assef2010}.
However, a spectrum taken after the transient emission fades is still required for a quantitative comparison to other TDE hosts and for disentangling the continuum level of the transient from the of the host. 

The host galaxy is seen face-on and has no obvious signature of spiral structure, apart from a faint hint of an extended structure to the West from the centre (Fig. \ref{fig:chart}).

%\subsection*{Flare spectrum}
\noindent
\textbf{Flare spectrum}. 
The flare spectrum, taken by PESSTO at $-3$ 
rest-frame days from the $I$-band maximum, supplemented with the earliest {\it Swift} observation in UVM2, taken at around the peak, is consistent with black-body model with temperature higher than 22,000~K. The host emission is clearly present in the spectrum at wavelengths longer than rest-frame 4000 {\AA}. The residual spectrum, after subtraction of the black-body continuum, shows two broad emission features around \Heii and H$\alpha$, as seen in all optically--selected TDEs. \cite{Arcavi2014} reported on a continuum of broad \Heii and H$\alpha$ emission lines (after the host-galaxy light has been subtracted) in the spectra of TDEs. 
OGLE16aaa seems to fit the picture very well, though with somewhat higher velocity dispersion ($\sim$23,000 km/s and $\sim$19,000 km/s for \Heii and H$\alpha$ lines, respectively, corresponding to FWHMs of $\sim$850{\AA}(54,000 km/s) and $\sim$970{\AA}(45,000 km/s), respectively). 

%sigma Gauss for HeII= 364, sigma Gauss Halpha=413; lambda_HeII = 4686; lambda_Ha = 6583; previous 18000 km/s -> sigma HeII = 280 -> FWHM = 660 A; 17000 km/s -> sigma Ha = 370 -> FHWM = 880 A

\noindent
\textbf{Light curve}.
There is no apparent flaring nor variability activity in the historical (3.5 yrs) OGLE $I$-band light curve of the nucleus of the host galaxy 
%(Fig. \ref{fig:history}) 
at a level below $I\sim$22 mag (host subtracted), less than 1 per cent of host light. 
The optical light curve reached the peak in about 30 rest-frame days and then exhibited significant variability, particularly around 15 restframe days after $I$-band maximum. 
Comparing our brightest UV measurements to the archival GALEX data we estimate the overall UV amplitude of the flare of about 3 mag.  
The overall decline of the light curve in both optical and UV is very slow, however, the actual slope of the decline is yet to be determined in the observations in the second half of 2016. 
No X-ray emission was detected by {\it Swift}/XRT at a level above 0.002 counts/s (3$\sigma$), corresponding to an upper limit for the unabsorbed luminosity of 5E42 erg/s (0.3-10 keV), for a power-law with photon index of -2.

\noindent
\textbf{Tidal Disruption Model}.
%Figure \ref{fig:tdemodel} shows the 1-$\sigma$ ensemble of light curves fits to the event produced by the {\tt TDEFit} code \citep{Guillochon14,Vinko15}. Since the event is still ongoing the highest-likelihood parameters we find are still preliminary, and matches the overall evolution of the presently available data well, but with a significant amount of variance ($\sim 0.15$ mag) relative to the mean flux at a given time. The posterior distributions from {\tt TDEFit} suggest a mass for the black hole of 
%$\text{Log}_{10} M_{\rm h} = 6.2\pm0.1$
%and mass for the disrupted star of 
%$\text{Log}_{10} M_\ast = -0.5\pm0.4$ ($\sim$0.3$\msun$). 
%
We fit OGLE16aaa's photometry with the tidal disruption light curve fitting software {\\TDEFit}\citep{Guillochon14,Vinko15} (see Fig. \ref{fig:tdemodel}), a Monte Carlo modelling code. For OGLE16aaa, we presume that the observed light comes from a combination of the light produced by a viscously-driven disk component (Guillochon et al. 2016, in prep)
, emission from circularisation\citep{Jiang16}, and reprocessing of light from the debris that ensheaths the accretion disk structure\citep{Guillochon14}. 
We assume a flat prior for $M_{\rm h}$ that allows all black hole masses between $10^4$ and $10^9$ $\msun$, a prior on impact parameter $\beta$ 
%$\equiv r_{\rm p}/r_{\rm t}$ 
that assumes pinhole scattering ($P(\beta)\propto \beta^{-2}$), a Kroupa stellar mass function
\citep{Kroupa2001}
, and flat priors on all other parameters. 
We find that the most likely combination of disruption parameters is
$M_{\rm h} = 10^{6.2\pm0.1} \msun$,
$M_{\ast} = 10^{-0.5\pm0.4} \msun$ (between 0.1 and 0.8 $\msun$, median $\sim$0.3$\msun$) 
and $\beta = 1.77_{-0.53}^{+0.94}$,
with a degeneracy between a sub-solar star suffering a full disruption and a solar star suffering a partial disruption.
The total observed energy emitted in the event is about 5e52 ergs. 
For assumed (median) 8 per cent of efficiency the total accreted mass is therefore about 0.3 $\msun$, indicating either complete or partial disruption of the star and suggesting the mass of the star was probably higher than 0.6 $\msun$.

% \begin{table}
%  \caption{TDE model parameters. See \citet{Guillochon14} for details on the parameters.}
%  \label{tab:tdefit}
%  \begin{tabular}{ll}
%   \hline
% $\text{Log}_{10} \mathcal{V}$ & $-1.94_{-1.28}^{1.32}$\\
% $a_{\text{spin}}$ & $0.502_{-0.353}^{0.309}$\\
% $\beta$ & $1.77_{-0.534}^{0.935}$\\
% $l$ & $0.904_{-0.382}^{0.579}$\\
% $\text{Log}_{10} f_{\rm out}$ & $-0.531_{-0.469}^{+0.341}$\\
% $\text{Log}_{10} M_{\rm h}$ & $6.23_{-0.0990}^{+0.108}$\\
% $\text{Log}_{10} M_\ast$ & $-0.498_{-0.356}^{+0.399}$\\
% $\text{Log}_{10} N_{\rm h}$ & $18.4_{-1.01}^{+1.20}$\\
% $\kappa  \text{Log}_{10}$ & $-0.220_{-0.139}^{0.393}$\\
% $\phi$ & $0.673_{-0.434}^{+0.257}$\\
% $\text{Log}_{10} R_{\text{ph}}$ & $-1.89_{-0.929}^{+0.976}$\\
% $\text{Log}_{10} R_{\text{sc}}$ & $-0.0676_{-0.109}^{+0.0527}$\\
% $R_v$ & $4.37_{-1.60}^{+1.16}$\\
% $t_{\text{off}}$ & $856000._{-373000.}^{+355000.}$\\
% $\text{Log}_{10} \sigma _v$ & $-0.848_{-0.0439}^{+0.0485}$\\
% $\text{Log}_{10} \tau _{\text{visc}}$ & $6.76_{-2.70}^{+1.25}$\\
%   \hline
%  \end{tabular}
% \end{table}

%%%%%%%%%%%%%%%%%%%%%%%%%%%%%%%%%%%%%%%%%%%%%%%%%%
\section{Discussion}
The characteristics of OGLE16aaa resemble those of other optically found Tidal Disruption Events. 
First, the flare's location coincides with the nucleus of the host galaxy.
Also, the derived photospheric black-body temperature remains high ($\sim$20,000~K) throughout the available data for this event, significantly higher than in typical supernovae. %and consistent with those observed in other TDEs.
Moreover, the temperature seems to rise at 35 rest-frame days from $I$-band maximum, as seen before in ASASSN-14ae TDE \citep{Holoien2014}.
Both the optical light curve and the presence of very broad \Heii and H$\alpha$ in the spectra resemble those of other optical/UV TDEs \citep{Arcavi2014}.
OGLE16aaa seems, therefore, to be a TDE. 
Lack of any variability in 3.5 years prior to the flare and large amplitude of the flare strongly disfavours regular AGN flaring. 

However, the underlying host galaxy is somewhat different in OGLE16aaa than in most of known TDEs so far. 
\cite{FrenchArcavi2016} have shown that most optically-found TDEs detected so far occur in 
quiescent, Balmer-strong galaxies, which are considered post-mergers. 
The Balmer absorption line series are not present in case of OGLE16aaa, however, a deep post-flare spectrum is still needed to verify it. Another TDE candidate, SDSS J0748, also does not show Balmer series and is an outlier on Figure 2 of \cite{FrenchArcavi2016}.

Among the X-ray-detected TDEs, IGR J12580, interpreted as due to a flare due to disruption of a Jupiter-mass planet was also detected in a weak AGN/Seyfert galaxy NGC4845 \citep{2013A&A...552A..75N}. 
The narrow emission lines ratios in all those three hosts, as well as their WISE colours indicate that the host contains weak-AGN.
%pgj ASASSN14li also has a pre-TDE AGN as shown by the detection of radio emission, see van Velzen et al. 15 Science.
% IA: Also Alexander et al. 2016.

For most TDEs found so far, the black hole was assumed dormant, since there was no evidence to the contrary.
Here we propose that OGLE16aaa and several other TDE candidates are due to a stellar disruption in a weak-AGN or Seyfert 1-type galaxy, where narrow emission lines originate from the circumnuclear material, photoionised by X-ray photons generated due to accretion. 
\citet{2006A&A...459...55B} showed that for several Seyfert II galaxies the projected distance of the narrow-line region extends to hundreds and even thousands of parsecs. 
Such accretion is likely to have been due to regular AGN accretion, but it could also have been due to a previous TDE. 
TDEs are expected to repeat on timescales of 10$^4$ years (\citealt{2004ApJ...600..149W}) and if the paths from the lines of sight from the narrow line region to us represent a broad delay function, we could still see an echo of previous TDEs in the spectra (\eg \citealt{2011ApJ...738L...8W}).
%light echoes of previous TDE flares seen in narrow emission lines, also Clausen D.: http://ciera.northwestern.edu/Aspen2015/abstracts/DrewClausen-poster.pdf

Moreover, the TDEs in narrow emission line hosts seem to be bridging stellar disruptions by dormant SMBHs and those in much more active AGN, which exhibit a transient appearance of broad-line \Heii and/or H$\alpha$ and increase in blue continuum in previously QSO-like spectra with strong narrow emission lines (\eg \citealt{Macleod15}). \cite{Merloni2015} has already suggested that the first example of CH-L-QSO from Stripe 82 can actually be a stellar disruption on a fairly massive SMBH ($\sim 10^8~\msun$). 
In such a case, the observed effect of a star getting too close to the main engine is the luminosity increase due to an increase in accretion rate (blue continuum from the thermal emission) and the disrupted material allows for the formation of broad \Heii and H$\alpha$ lines with very large dispersion. 
If TDEs are common enough, especially in post-merger galaxies, where the disruption rate is increased either due to dynamical instabilities of the nuclear cluster, or a presence of a binary black hole, they could play an important role in growth of nuclear black holes (\eg \citealt{Hills1975}, \citealt{2007ApJ...667..704V}).

%%%% variability pre-flare due to AGN
While AGN are known to vary on small scales, none of the hosts of weak AGN TDEs exhibit any significant pre-flare photometric variability. Primarily, this is due to lack of long-term photometric data for most of them. OGLE16aaa is probably the first TDE candidate in a weak AGN where we have a long (3.5yrs) history of pre-flare observations indicating no detectable variability. 

%%% rates - bias
Based on pre-flare variability, \cite{vanVelzen2011} rejected several candidates in Stripe 82 data, and since then, all the current surveys are using this criterion when selecting TDEs among nuclear flares. 
However, as already suggested in \cite{Strotjohann2016}, significant and temporal (flare-like) changes in the accretion in AGN are hard to explain with changes in gas flow and can be attributed to disruptions of stars.
%(see also \citealt{Torricelli-Ciamponi2000}, \citealt{Kennedy2016}). 
%(\eg \citealt{Kennedy2016}). 
Restricting nuclear flares solely to apparently quiet nuclei might introduce biases in observational optically-selected TDE rate determinations.

The $\sim$monthly variability observed in the optical and UV bands during the flare and seen in the residuals of the TDE model, remains puzzling in OGLE16aaa. 
Possible explanations include precession of the disk (\eg \citealt{Tchekhovskoy14, 2011MNRAS.414.2186J}), duty cycle imposed by the orbital period of the returning debris\citep{Jiang16}, or even a binary SMBH\citep{2014ApJ...786..103L}.
%
%Extreme X-ray light curve variability has been observed in Swift J1644 event 
%(\eg 
%%\citealt{Saxton2012}, 
%\citealt{2014ApJ...784...87S}) with quasi-periodic oscillations with timescales evolving from several hours to about a month, possibly due to modulation of the jet luminosity by the precession of the disk \citep{Tchekhovskoy14}. 
%The precession of the disk can also cause the mass accretion rate to vary \citep{2011MNRAS.414.2186J}, hence the variability is also expected to be seen in the thermal emission. 
%Another potential source of variability is the duty cycle imposed by the orbital period of the returning debris 
%%\citep{Guillochon15,Jiang16}, 
%\citep{Jiang16}, 
%which depending on how relativistic the encounter is can yield variations on timescales from days to years. 
%A star disrupted about a binary SMBH can also yield extreme variability \citep{2014ApJ...786..103L}. 
Further observations are required to address this issue.

%%% variability -> Binary black hole
% The variability of the photometric light curve of OGLE16aaa remains a mystery. If the modulation remains periodic at later epochs, as seen so far with a period of about 27 rest-frame days, one possible explanation could be a presence of a binary black hole. 
% \cite{2014ApJ...786..103L} have found the first case of tidal disruption exhibiting signatures of a binary black hole in SDSS J120136.02+300305.5 with a period of 150 days seen in X-rays. Their model suggested a binary black hole at separation of 0.6 mpc. In case of OGLE16aaa the separation would be even smaller, around 0.2 mpc, making this the tightest black hole binary among the supermassive ones. Further  observations are required to confirm the periodicity. 

\section{Conclusions}
In the course of OGLE-IV search for extragalactic transients, we have discovered a new candidate for a TDE of a 0.1-0.8$\msun$ star by a $10^{6.2}\msun$ SMBH.
%in the very centre of a galaxy at z=0.1655. The characteristics of the transient are in agreement with a Tidal Disruption Event caused by a central low-mass black hole (log($M_{BH}$)$\approx$6.5), disrupting a low-mass main sequence star. 
%Spectrum at around maximum optical brightness shows signatures of broad \Heii and H$\alpha$ lines, due to the disk of debris formed around the black hole. 

%The host of OGLE16aaa is most likely a weak-AGN galaxy, which is unusual for TDEs found so far. 
OGLE16aaa event, along with 
SDSS J095209.56+214313.3,
SDSS J0748,
ASASSN-14li
and IGR J1258
seem to form a sub-class of TDEs in galaxies hosting a weak AGN, with weak narrow emission lines. Moreover, the existence of such TDEs supports one of the explanations for so called ``changing-Look QSOs'', where persistent strong narrow emission lines get super-imposed with variable broad emission lines. 

A possible explanation of the observed variability in the light curve of OGLE16aaa is that it is induced by a binary black hole on a tight orbit or due do disk precession or circularisation on a timescale of about a month, however, further multi-messenger follow-up is required to understand this. 

The fact that TDEs could also be found in (weak) AGNs is important for determining TDE rates, since currently there is likely an observational bias against selection of optically found TDEs. 
Whereas only 10 per cent of galaxies host an AGN this bias could be larger given that AGN activity is triggered or enhanced by recent mergers and milliparsec-scale binary SMBHs could strongly enhance the TDE rate.

%%%%%%%%%%%%%%%%%%%%%%%%%%%%%%%%%%%%%%%%%%%%%%%%%%
\section*{Acknowledgements}
We would like to thank Drs. Marek Niko{\l}ajuk, Dan Maoz, Jozsef Vinko and Sjoert van Velzen, and the attendees of the Warsaw Black Hole Lunches, for helpful discussions.
OGLE Team thanks Profs. M. Kubiak and G. Pietrzy{\'n}ski.

{\L}W, AH and KR were supported by Polish National Science Centre grant OPUS 2015/17/B/ST9/03167. 
PGJ and ZKR acknowledge support from ERC Consolidator Grant 647208. 
OGLE project has received funding from the Polish NCN, grant MAESTRO 2014/14/A/ST9/00121 to AU.
IA is partially supported by the Karp Discovery Award.
%TK acknowledges support through the Sofja Kovalevskaja Award to P. Schady.% from the Alexander von Humboldt Foundation of Germany.
This work is based (in part) on ESO PESSTO programme 188.D-3003, 191.D-0935. SJS acknowledges ERC grant 291222 and STFC grants ST/I001123/1 and ST/L000709/1.
We acknowledge the use of NED database (ned.ipac.caltech.edu) and TDE catalogue (tde.space).


% Don't change these lines
%\bsp	% typesetting comment
\label{lastpage}
\end{document}